\documentclass[aps,twocolumn,pre,showpacs]{revtex4}

\usepackage{latexsym}
\usepackage[pdftex]{graphicx}
\usepackage{graphicx}
\usepackage{epstopdf}
\newcommand{\la}{\langle}
\newcommand{\ra}{\rangle}

\begin{document}

\title{Optimization of transport protocols with path-length 
constraints in complex networks}

\author{Jos\'e J. Ramasco}\email{jramasco@isi.it}
\affiliation{Complex Systems Lagrange Laboratory, (CNLL), 
ISI Foundation, 10133 Turin, Italy}

\author{Marta S. de La Lama}\email{msanchez@ifca.unican.es}
\affiliation{Max-Planck Institute for Dynamics and Self-Organization. Bunsenstr. 10, 37073 G\"ottingen, Germany}
\affiliation{Instituto de F{\'i}sica de Cantabria (CSIC-UC), Av. de los Castros 
s/n, 39005 Santander, Spain}
\affiliation{Departamento de F\'isica Moderna, Universidad de Cantabria,  Santander, Spain}

\author{Eduardo L\'opez}\email{eduardo.lopez@physics.ox.ac.uk}
\affiliation{CABDyN Complexity Centre, Sa\"{\i}d Business School, University of Oxford, Park End, Oxford OX1 1HP, UK}
\affiliation{Department of Physics, University of Oxford, Parks Road, Oxford OX1 3PU, UK}

\author{Stefan Boettcher}\email{sboettc@emory.edu}
\affiliation{Physics Department, Emory University, Atlanta Georgia 
30322, USA.}

\date{\today}

\begin{abstract}

We propose a protocol optimization technique that is applicable to both
weighted or unweighted graphs. Our aim is 
to explore by how much a small variation around the Shortest Path or Optimal Path protocols can enhance
protocol performance. Such an optimization strategy can be necessary because even though some protocols can achieve very high traffic tolerance levels, this is commonly done by enlarging the path-lengths, which may jeopardize scalability. We use ideas borrowed from Extremal Optimization to guide our algorithm, which proves to be an effective technique. Our method exploits the
degeneracy of the paths or their close-weight alternatives, which significantly 
improves the scalability of the protocols in comparison to Shortest Paths or Optimal Paths protocols, keeping at the same time
almost intact the length or weight of the paths. This characteristic ensures that the optimized routing protocols are composed of paths that are quick to traverse, avoiding
negative effects in data communication due to path-length increases that can become specially relevant when information losses are present.
\end{abstract} 
\pacs{89.75.Hc, 89.20.Hh, 05.70.Ln} 
\maketitle

\section{Introduction}
Communication systems are an essential element of the modern world, and represent some of the most noticeable settings of transport problems in complex structures. The Internet, telephone systems, power grids, etc., provide ways to transport information, signals or resources from one part of the globe to another. Such communication is done by means of highly structured systems that can be modeled by networks, {\it i.e.}, collections of elements corresponding to nodes, links that interconnect them, and weights on the links symbolizing connection  capabilities~\cite{albertrev,newmanrev,sergeibook,romubook,alainbook}.

In many examples of communication systems, discrete subsets of information (packets) are sent from a source to the appropriate destination. Typically, a multitude of users simultaneously  perform
similar tasks between many different  source-destinations pairs, and thus the network at any given time is populated by numerous traveling data packets. In order for packets to find their way, routing protocols are put in place which determine how the packets travel from source to destination. Optimization of the performance in communication systems can thus be achieved in two non-exclusive ways: improving the structure of transport networks~\cite{guimera02}, and finding optimal routing protocols~\cite{ohira98}. Both approaches have been the focus of a great deal of research on this subject in recent years~\cite{guimera02,ohira98,arenas01,sole01,echenique04,echenique05,lopez06,braunstein07,Cohen,LPP,braunstein03,goh01,marc03,sree04,sree06,danila06,danila07,yan00,kalisky05,goh05}.  A number of important results have been obtained, elucidating the usefulness of so-called scale-free networks~\cite{guimera02,lopez06}, the vulnerability of networks to random or targeted failures of some of its elements~\cite{Cohen}, and the relevance of path-lengths in the efficiency of communication~\cite{LPP}.

When the overall traffic in the network is very large, the system may get compromised  in its capability to deliver  information. The origin of this problem could be twofold: either the 
connections of the network are unable to tolerate such a high traffic flow or, more commonly, 
some components (nodes) get overloaded, {\it i.e.}, congested.  Congestion occurs when the nodes 
become heavily used acting as bottlenecks for the network communication. 
By establishing clever routing protocols, this congestion can be diminished or, for 
moderate rates of overall traffic in the system, eliminated. The assessment of node work-load 
requires thus the introduction of a new quantity.  Betweenness centrality (betweenness in short) has 
been proposed as such a measure~\cite{guimera02,goh01}, as an estimate of how much use a given node can expect to receive. 
Specifically, for a set of paths in a network, the betweenness of a node is defined as the number 
of paths visiting that node. Therefore, a given routing protocol induces a betweenness on the nodes of the network.
It has been pointed out that nodes with large betweenness are the first to form packet queues
when traffic is large enough~\cite{guimera02}. 
If this queue is not eliminated, then the travel times of 
packets from source to destination increase without bound, hence, rendering the node (and potentially the network) congested. 

On the other hand, packet delivery is an imperfect process since it can be subjected to
packet loss or corruption. When congestion occurs, even entire
queues of packets could be eliminated while waiting at overloaded nodes. Thus, when attempting
to optimize a routing protocol, care must be taken not to increase the path-length
to a level that would elevate the likelihood of packet loss beyond 
critical limits. Such considerations have led to the formulation
of models such as Limited Path Percolation, which exhibit a phase transition
when the length of paths increases beyond a specific tolerance level~\cite{LPP}. 

In this article, we focus on the problem of optimizing the transport rules 
for scale-free networks with properties
similar to those of computer networks. However, in contrast to a number of other studies, we
optimize the protocols to reduce the likelihood of
congestion {\it but} keeping path-lengths as close as possible to the shortest, thus guaranteeing efficient
communication paths. We find that by applying ideas borrowed from Extremal Optimization (EO)~\cite{boett01}, 
which uses local information to approach the optimal state of a system, we can achieve both
goals of congestion avoidance and short path-lengths.

The article is structured as follows: Sec.~\ref{algorithm} explains the philosophy behind our
optimization heuristic, the algorithm to find
optimal routing protocols, and our network model. Sec.~\ref{betweenness} quantifies
betweenness of nodes as a consequence of the optimized routing protocol. 
Sec.~\ref{paths} focuses on the path-lengths created by the new, EO-based, routing protocol, and the advantages
that it affords regarding maintaining the path-lengths within a constant factor of the original paths.
Sec.~\ref{weights} extends our methods to weighted networks, and finally, Sec.~\ref{conclusions}
is dedicated to the conclusions.

\section{Optimization method}
\label{algorithm}

To define the betweenness precisely, 
we need first to specify a routing protocol, which is 
a set of paths $\mathcal{L}=\{\rho_{ij}\}$ between all pairs of nodes $i$ and $j$
of the network. We define the length of each path
$\ell_{ij}$ as the minimum number of links needed to be crossed when going from $i$ to $j$ following the path $\rho_{ij}$.
We adhere to the convention of considering only one path 
for each pair of nodes~\cite{sree06}, as the length degeneracy of the 
paths will be exploited by the method to 
optimize the protocol.  The rules by which these paths are chosen can be numerous.
For instance, one typical choice for $\mathcal {L}$ in the literature is the set of shortest paths between
nodes (SP). 
The betweenness $b_i$ of a node $i$ is then defined as the number of paths of $\mathcal{L}$ that visit 
node $i$, without counting those paths that begin or end in $i$. (This choice is somewhat arbitrary,
and other choices are also possible.) Note that in the literature the word 
"betweenness" sometimes is associated exclusively with the number of paths crossing a node within the framework of the SP 
protocol. We are using this concept in a more general context since it can refer to any routing protocol, 
which has to be given. In fact, betweenness values depend
on both, the structure of the network and $\mathcal{L}$. For a given network,
changing the routing protocol typically changes ${b_i}$ over the network except in very
special cases such as when there is only one possible protocol choice (e.g., a star network).

Given a network and a protocol $\mathcal{L}$, it has been shown that the first
node to begin accumulating a queue, say $m$, satisfies that
$b_m>b_i$ for all nodes $i\neq m$ in the network~\cite{guimera02}. To
explain this point, consider a network of size $N$ in which data packets are produced in each node at 
random with a rate $0\leq\gamma\leq 1$. In the simplest version of the problem, the destination of each 
packet is random too, and thus, each node can produce a packet to any of the other $N-1$ nodes of the network 
with equal probability. In this case, it was found that the number of packets in
node $i$ is, on average, $\gamma\,  b_i/(N-1)$~\cite{guimera02}. Without loss of generality, we assume that 
each node can processes (route through) a single packet per unit time, and thus, the first node to 
congest is the one with the maximum betweenness value, $b_{\rm max}$. We can also rephrase this point 
by indicating that the critical rate of packet production would be $\gamma_c=(N-1)/b_{\rm max}$ before 
some node of the network congests~\cite{guimera02}. This relation shows the relevance of designing 
routing protocols with a $b_{\rm max}$ as low as possible in order to prevent network congestion for low 
$\gamma$ values. The lower $b_{\rm max}$ is, the higher the number of packets can be (higher $\gamma$) 
before the network congests. There is, however, a price to pay in this strategy because the reduction of $b_{\rm max}$ 
is usually attained at the expense of enlarging the path-lengths. Under some circumstances, such as when packets 
can get lost, enlarging significantly the paths may not be the best option. The model that we propose 
next aims at bridging the gap between these two opposing forces: we explore how much $b_{\rm max}$ can be reduced 
while keeping the paths as close to their shortest version as possible.



We generate the networks for the analysis with the configuration model~ \cite{reed-molloy95,romu05} that allows us to build graphs with arbitrary
degree sequence but free from degree-degree correlations. The degree of a node $i$ corresponds to its number of connections and is usually represented as $k_i$. In this work, we
will focus on power-law degree distributed networks, $P(k) \sim k^{-\lambda}$,
with exponent $\lambda = 2.5$ to facilitate the comparison with previous 
studies~\cite{sree06,danila06,yan00}. Nevertheless, we have also performed the
analysis on networks with $\lambda = 4.5$ finding no qualitative difference in the final results.

The optimization algorithm starts from the SP protocol: Once the
network is built, the first step is to calculate the shortest paths between
all nodes of the network using Dijkstra's
algorithm~\cite{cormen}. 
Then, in an iterative way for as long as desired, we repeat the following steps:

\begin{itemize}
\item The node with the highest betweenness $m$ is selected.
\item One of the paths, $\rho_{ij}$, passing through $m$ is chosen at random; we refer to its initial and
final nodes as $i$ and $j$ (with $i \neq m \neq j$).
\item A path  between $i$ and $j$ is searched. The new path must be the shortest alternative to $\rho_{ij}$ not passing through $m$. In case that there are more than one alternative paths degenerate in length, one of them is selected at random. On the other hand, if 
there is no alternative to $\rho_{ij}$, the path remains without change.
\end{itemize}
\begin{figure}
\begin{center}
\includegraphics[width=8.6cm]{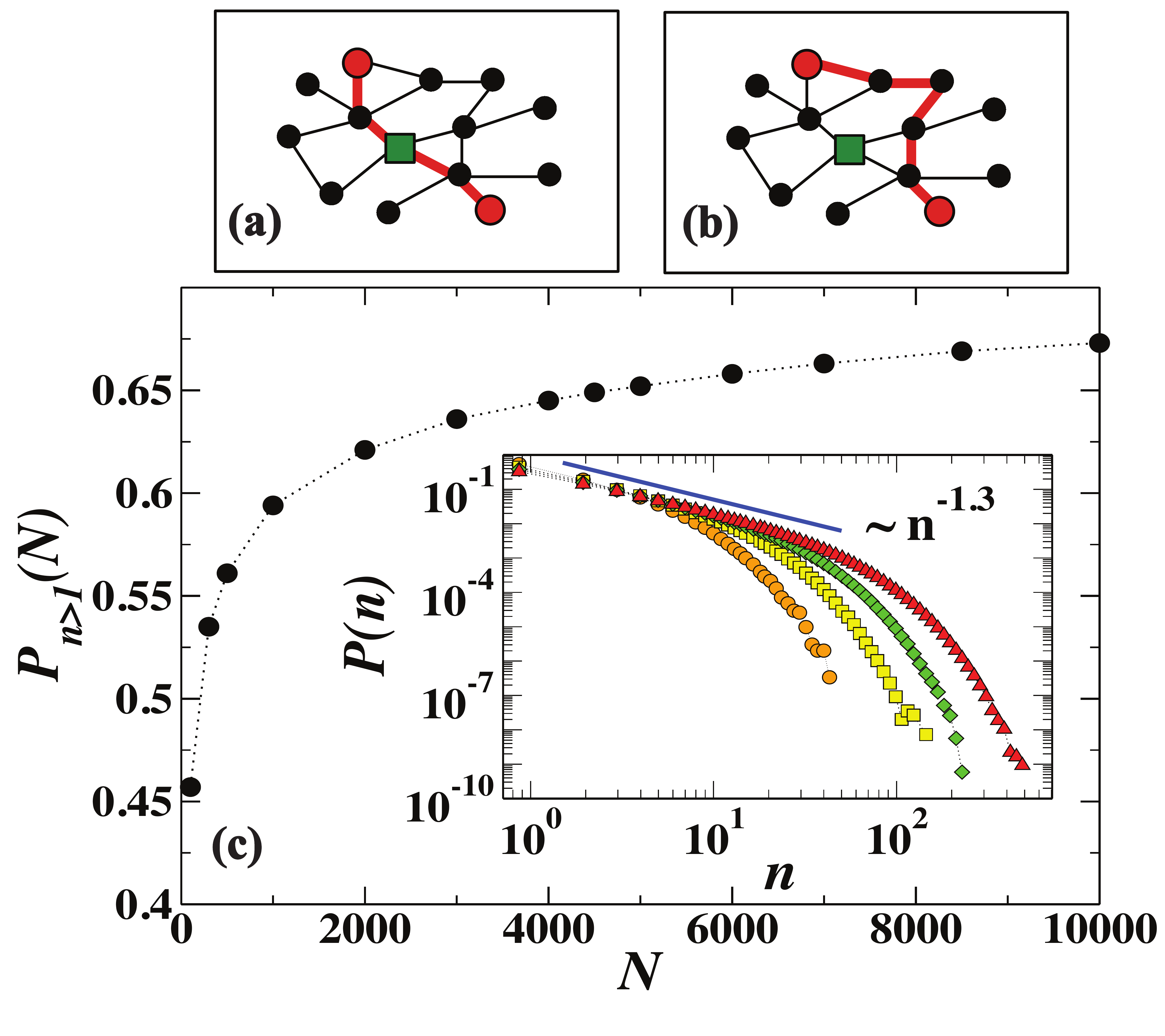}
\caption{(Color Online) The sketches (a) and (b) illustrate the way in which the method
optimizes the protocol by selecting another path that must be either of 
the same length or the shortest alternative to the original one. In panel (c), the
probability of finding alternative length-degenerate paths, $P_{n>1}$, for 
the
Shortest-Path protocol is displayed as a function of the 
network size $N$. In the inset: histogram for the 
number $n$ 
of degenerate alternatives for each path in the SP protocol. The different
curves correspond to the following network sizes: $N=100$ (circles), $500$ 
(squares), $2000$ (diamonds) and $10000$ (triangles).}
\label{fig:1}
\end{center}
\end{figure}
A schematic illustration of these rules is shown in panels (a) and (b) of 
Figure~\ref{fig:1}. In the sketch, a path passing through the highest betweenness
node, the square, is re-routed thereby reducing betweenness. It is 
worth stressing that, the constraint on the length of the alternative paths in 
practice means that if another choice of equal length to the original path exists, 
the path does not increase in length. The method takes advantage of the
possible degeneracy of the paths with respect to length to change the protocol and ease
the stress on the node(s) with the largest traffic.  Note also that the method keeps no long-term memory, the search for alternative paths due to sequential expulsions of a path from different nodes do not accumulate. The algorithm generates thus protocols with paths 
limited to SP configurations or the shortest
alternatives to them avoiding a single high used node.

The potential utility of exploiting the SP degeneracy and its importance in our networks can be seen in Figure~\ref{fig:1}. We have plotted in the panel (c) the probability of finding a degenerate alternative to the shortest path between any pair of nodes $i$ and $j$, $P_{n>1}(N)$, as a function of the network size $N$.   
$P_{n>1}(N)$ consistently grows with $N$, becoming higher than $65 \%$ for the largest graphs that we have considered 
$N = 10000$. In the inset, the distribution of the number of different degenerate paths, $P(n)$, is also shown for four network sizes. Note that $P_{n>1} = \sum_{n = 2}^\infty P(n)$.

The philosophy of our method is in the spirit of Extremal
Optimization (EO)~\cite{boett01}, attacking the worst element of the system 
in an attempt to improve the global performance. In the following, we will
study how its application affects the different features of the protocols.

\begin{figure}[b]
\begin{center}
\includegraphics[width=8.cm]{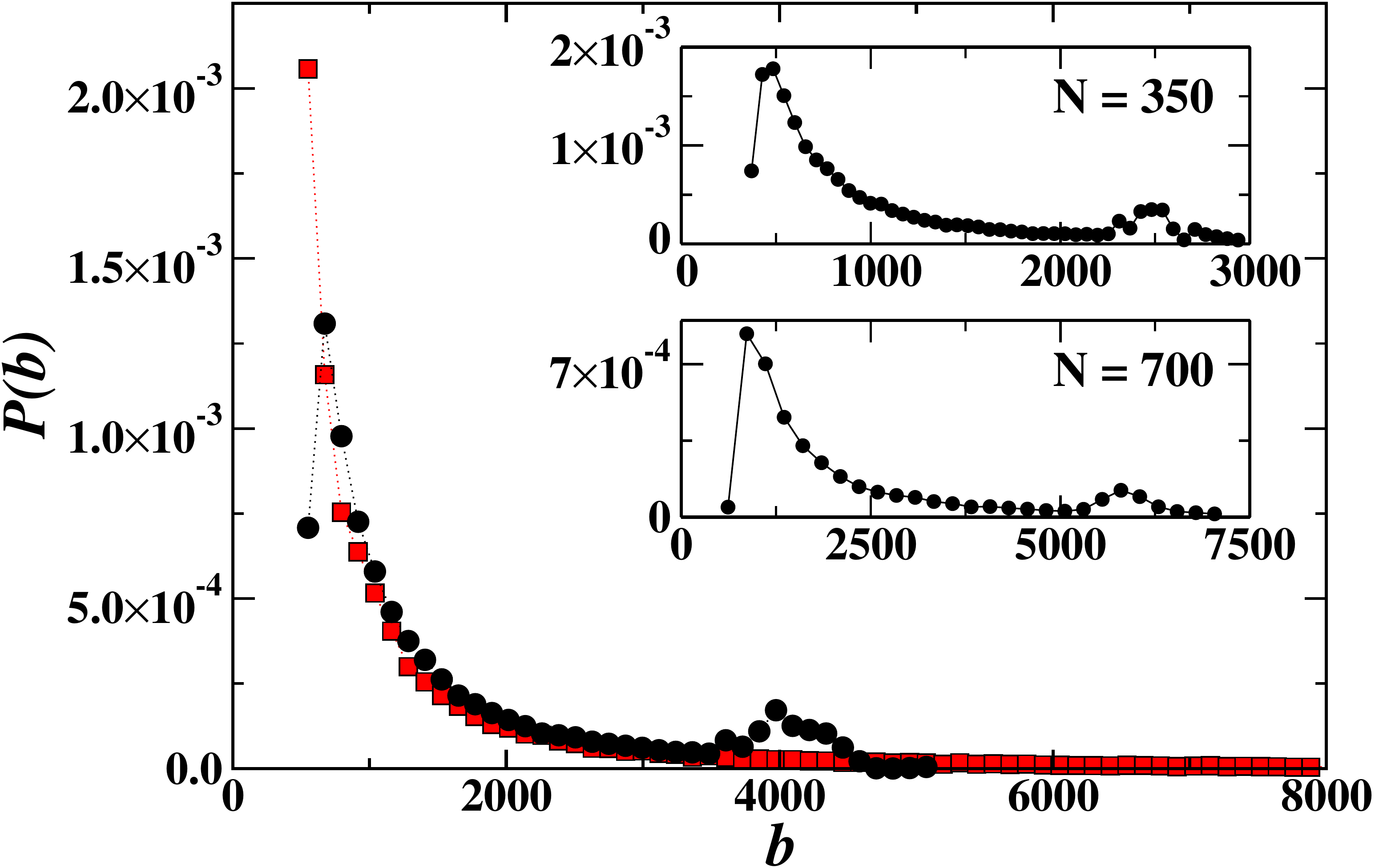}
\caption{(Color Online) Betweenness distribution for a $N=500$ system. The 
(black) circles 
correspond to the stationary distribution reached with our optimization method 
while the (red) squares are for the initial SP protocol. In the inset, the 
stationary betweenness distributions of the method for other two network 
sizes: $N=350$ and $N = 700$.}
\label{fig:2}
\end{center}
\end{figure}

\section{Betweenness distribution}
\label{betweenness}
We will discuss first the results concerning the betweenness of the
nodes and how its values adapt after the optimization method is employed. After some
iterations of the method~\cite{stat} the betweenness distribution, $P(b)$, reaches a 
stationary form that can be seen in Figure~\ref{fig:2}. Before describing it,
it is important to 
remember that the SP protocol produces a power-law distribution 
for $P(b)$ whose exponent depends on the exponent $\lambda$ of the degree
distribution \cite{goh01,marc03} (see the curve of (red) squares in 
Figure~\ref{fig:2}). The
protocols generated by the 
optimization method show that the tail of $P(b)$ collapses into a peak at 
high values of the betweenness. Also the area of low $b$ suffers a slight
variation, although the functional form of  $P(b)$ at intermediate $b$
seems largely unaffected. A change in the size of the network displaces 
the position of the peak but not the quality of the effects observed in the
main plot of Figure~\ref{fig:2}. 

\begin{figure}
\begin{center}
\includegraphics[width=8.cm]{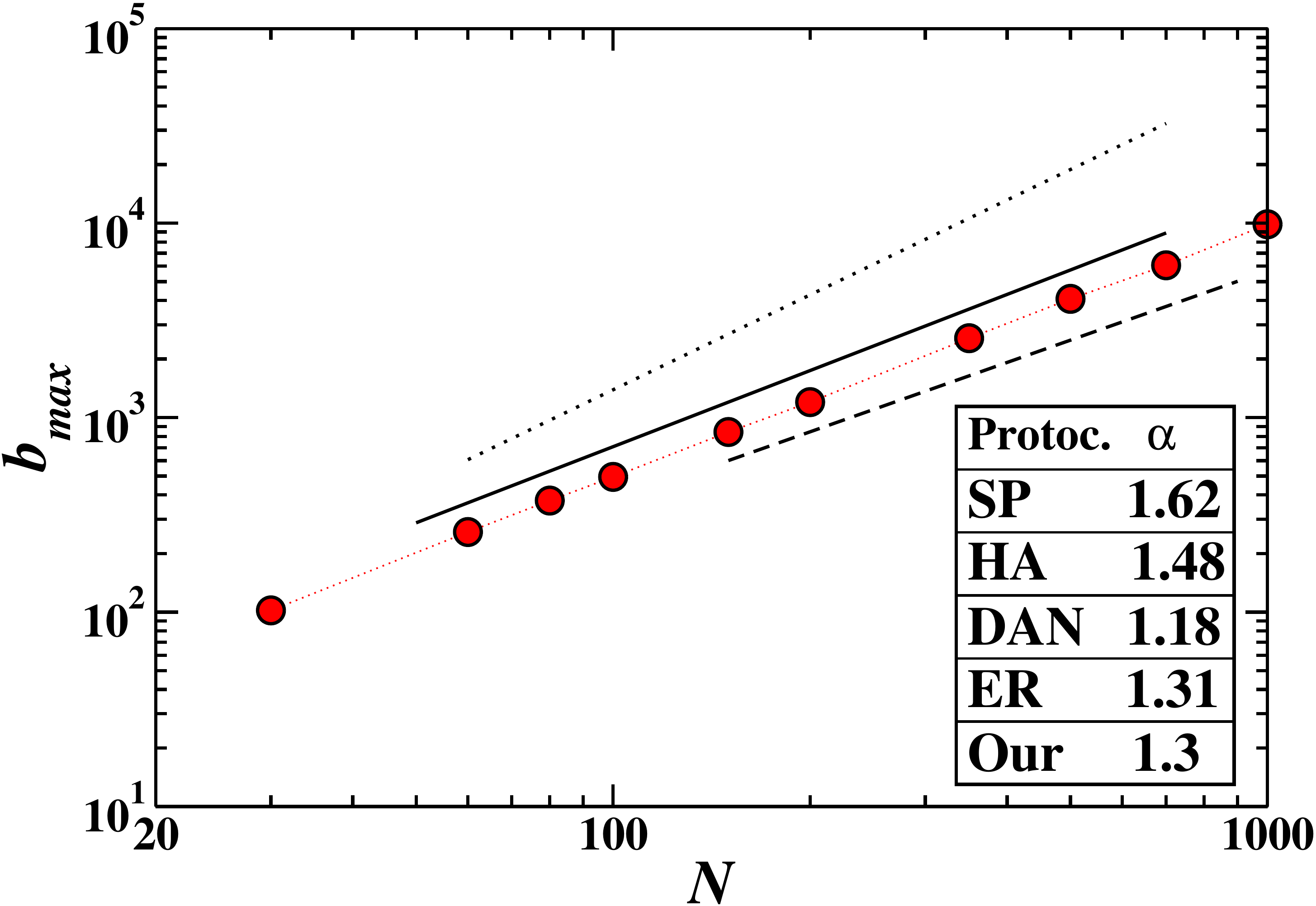}
\caption{(Color Online) The average maximum betweenness, $b_{max}$, as a function of the
network size $N$ for the optimized protocols. Lines are shown as a 
reference scaling with exponent $1.3$ (solid), $1.18$ (dashed), 
and $1.62$ (dotted). In the table, scaling exponents
$\alpha$ obtained with different optimization methods are listed. The initials
correspond to  
Shortest-Paths (SP)~\cite{yu07}, Hub-Avoidance (HA)~\cite{sree06}, Danila's method (DAN)~\cite{danila06}, 
Efficient Routing (ER)~\cite{yan00} , and our own method. 
}
\label{fig:3}
\end{center}
\end{figure}

The peak induced in $P(b)$ by the optimization method at high values of 
$b$ correlates with a decline of the maximum betweenness in the network. As mentioned before, this 
betweenness, $b_{max}$, is of crucial importance for the transport capability 
of a protocol, since it imposes an upper cutoff to the traffic that the network is 
able to sustain before congesting~\cite{guimera02,braunstein07}. The
scaling of $b_{max}$ with the network size, $N$, can thus be seen
as a measure of the scalability of a routing protocol: the faster $b_{max}$ grows
with $N$, the less useful the protocol is for large networks. In
Figure~\ref{fig:3}, the average $b_{max}$ in the stationary state of the
protocols obtained with our optimization method is depicted as a function of
the number of nodes in the network. We find a power-law increase 
with a functional form $b_{max} \sim N^\alpha$ and 
$\alpha^{our} \approx 1.3$. This value of $\alpha$ must be compared with the 
results encountered with other
protocols or other protocol optimization methods. In the same Figure, we have
also included a table with the exponents $\alpha$ for other methods. It is
specially interesting to compare with the original Shortest-Path protocol
$\alpha^{SP} \approx 1.62$~\cite{yu07} or with the best of the optimization
methods listed (Danila's method) with $\alpha^{DAN} \approx 1.18$~\cite{danila06,
danila07}. Our
optimization method, without deviating substantially from the SP protocol in
the length of the paths, 
produces an acceptable value of the exponent $\alpha$.

\section{Path-Lengths}
\label{paths}

\begin{figure}
\begin{center}
\includegraphics[width=8.cm]{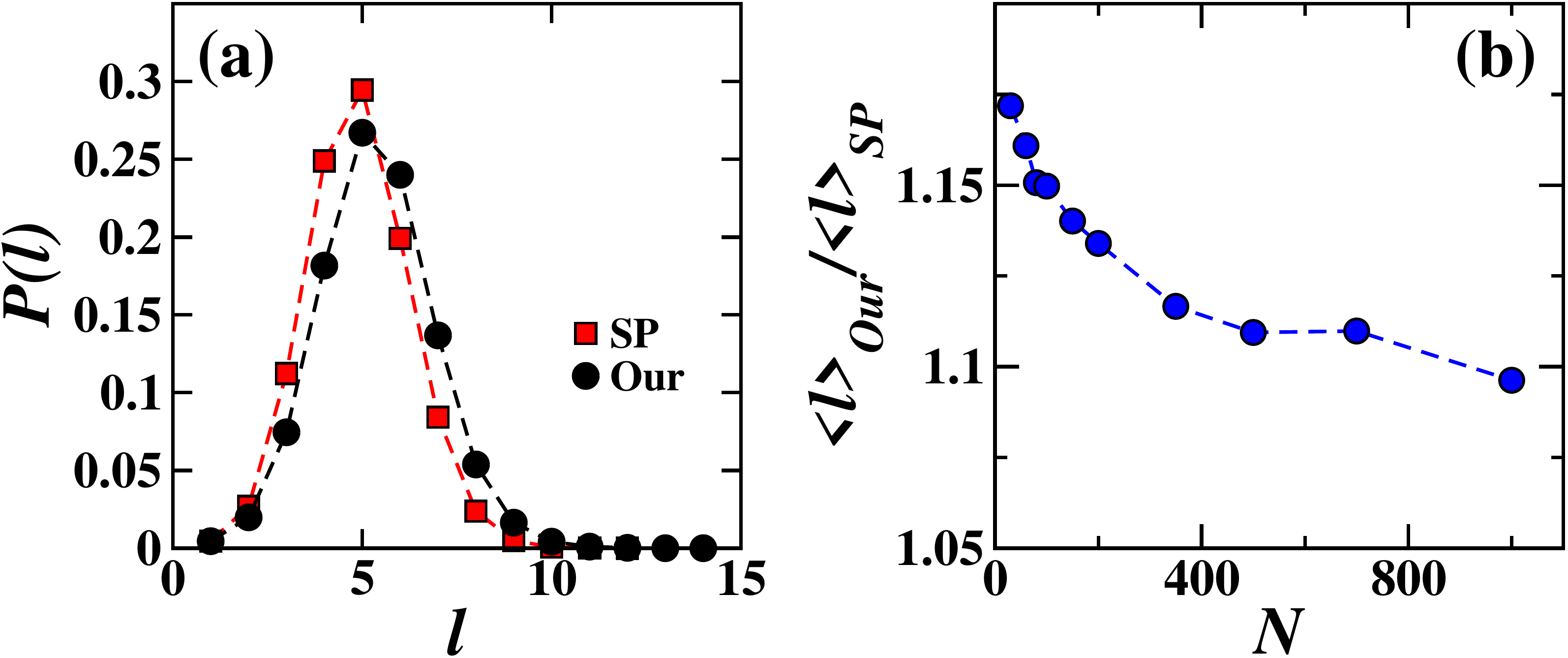}
\caption{(Color Online) In panel (a), path-length distribution for a network of size $N=1000$. In the
figure, the results of the protocols obtained with our optimization method can 
be compared with the initial SP. In panel (b), the ratio between the average length of
the paths in the two protocols (ours and SP) is visualized as a function of the
network size.}
\label{fig:4}
\end{center}
\end{figure}

The near-invariance of the path-length  between SP and our optimized protocol, and the
path-length distribution itself,  deserves further attention.
Here we will focus on studying how the length of the paths in the protocol 
mutates when the optimization method is applied to the protocol. In 
Figure~\ref{fig:4}, we have plotted the path-length distribution $P(\ell)$ for the
original Shortest Path protocol (red squares) and for the stationary regime of
the protocols obtained with our optimization method on a network of size $N =
1000$. As can be seen, the distribution is slightly displaced to larger values of $\ell$
but its shape remains essentially invariant. In order to quantify the relevance of
the change of $P(\ell)$ with the system size, we have represented
in panel (b) of Figure~\ref{fig:4} the ratio between the average length of the
paths in the protocols obtained with our method and that calculated with the SP. The curve
is monotonically decreasing with $N$, moving very slowly towards unity (or a
value just above). 

\begin{figure}[b]
\begin{center}
\includegraphics[width=8.cm]{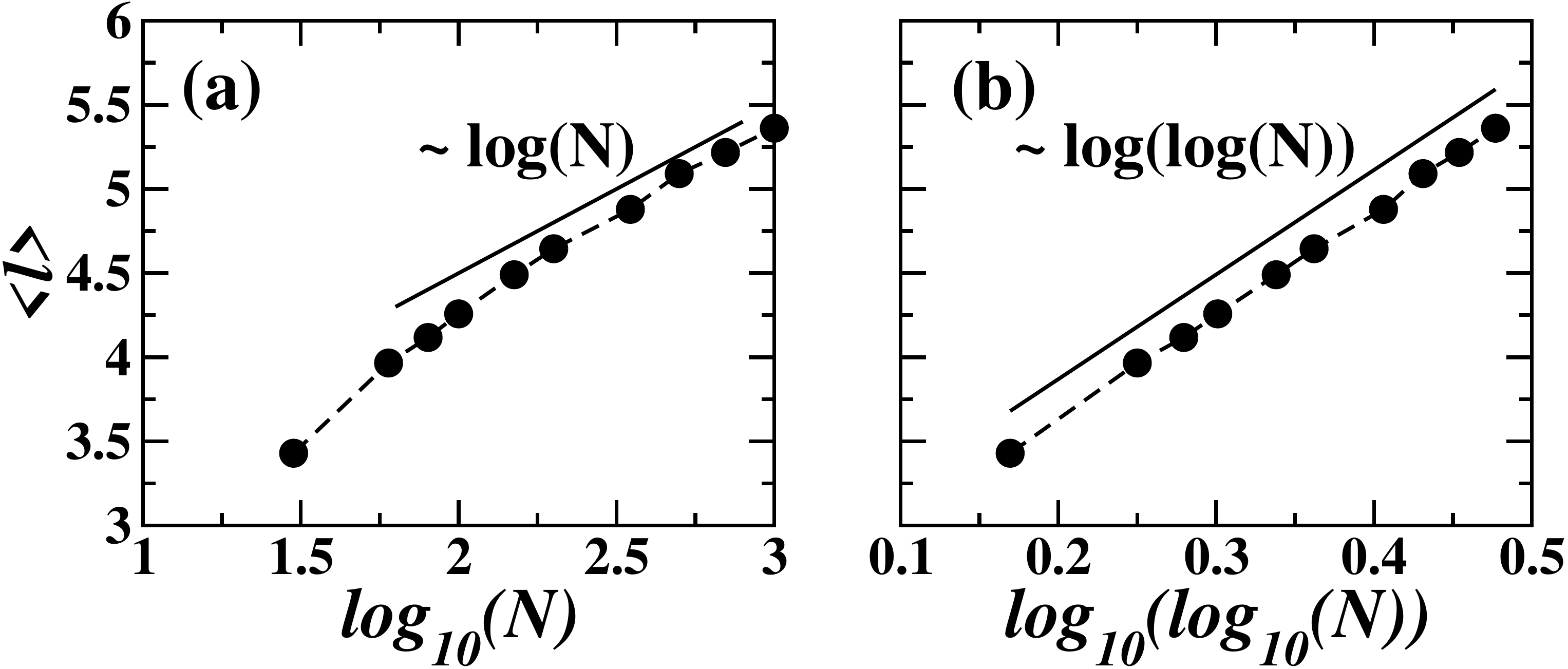}
\caption{Average path-length, $\langle l \rangle$, of the protocols we 
obtain as a function of the network size. It can be seen
that  a $log(log())$-scale in panel (b) provides a better fit to our simulation results
than a pure logarithmic scale in (a).}
\label{fig:5}
\end{center}
\end{figure}

Similar to $b_{max}$, limiting the scaling of the average length of the paths, 
$\la \ell\ra$, with the system size is also an
important design feature. A protocol that elongates the paths unnecessarily may not be efficient for
network communications. Generally, there exists a nonzero probability of losing
a packet in every communication between two servers, and the longer the paths become,
the higher the probability of missing information. Furthermore, if data loss 
reaches a point in which most of the paths are not functioning, the network suffers a
disruption similar to a percolation transition. In fact, this phenomenon has
been studied recently  and it is known as Limited Path Percolation~\cite{LPP}. 
In Figure~\ref{fig:5}, we show how $\la \ell\ra$ in the protocols obtained using our method behaves with increasing network sizes. In scale-free
networks with $\lambda = 2.5$, the SP protocol $\la \ell_{SP}\ra$ is expected to
grow as $\log({\log({N})})$. We cannot numerically test this formula for many
decades in $N$ for our optimized protocols, but within the limited range of values that we can explore the relation holds. Such scaling is important because, as mentioned above, the optimization of the protocols can in some cases imply a trade-off between reductions in $b_{max}$ and escalating path-lengths. For instance, Danila's method~\cite{danila06}, which exhibits the best scaling for $b_{max}$ versus $N$ of the methods shown in Figure~\ref{fig:3}, leads to a $\la \ell\ra \sim \log({N})$~\cite{danila07}, considerably faster than the $\log({\log({N})})$-scaling of our method (or of the SP protocol). This difference in the scaling can be relevant if the packet communication has losses in long paths (or in paths longer than the SP).

In general, consider a protocol that is tuned to scale
with the typical structural path lengths of the network, which, to be concrete, we associate
with the most probable length $\ell_{SP}^*$ of the length distribution $P_{SP}(\ell)$. 
We imagine that the real system uses a
protocol $\mathcal{L}_o$ when uncongested which is, for our purposes, the best practical protocol
for the network of interest (even if it is not strictly optimal). 
For a power-law network with exponent $\lambda=2.5$, 
$\ell^*_o=D_o\ell^*_{SP} \sim D_o\log\log(N)$, where $D_o\geq 1$ 
is a numerical constant. Now, when higher rates of packet creation appear on the network, to avoid congestion
we consider introducing two new protocols $\mathcal{L}_S$ and $\mathcal{L}_L$,
the first one with typical path lengths $\ell^*_S\sim \delta_S\ell^*_o$ ($\delta_S>1$ another constant), 
and the second with $\ell^*_L=F(\ell^*_o)$, with $F(x)>x^{1+\epsilon}$, $\epsilon>0$. In the thermodynamic limit,
with mean field approximation, $\tau$, the ratio of new typical path length to old typical path length in $\mathcal{L}_S$ 
remains constant, as we can see from
\begin{equation}
\tau_S=\frac{\ell^*_S}{\ell^*_o}=\frac{\delta_S D_o\log\log(N)}{D_o\log\log(N)}\rightarrow\delta_S.
\end{equation}
On the other hand, for $\mathcal{L}_L$, we find
\begin{equation}
\tau_L=\frac{F(D_o \log\log(N))}{D_o\log\log(N)}\rightarrow (D_o\log\log(N))^\epsilon\rightarrow\infty
\end{equation}
where the last relation occurs for $N\rightarrow\infty$.
In Ref.~\cite{LPP}, the limited path percolation transition was shown 
to be intimately related to $\tau$. 
For our purposes, it means that if a protocol keeps $\tau$ finite as $N\rightarrow\infty$, occurring 
in the $\tau_S$ case, there is
the possibility to have a protocol that can be tuned to maintain at least
a level of communication between node pairs. Note that
the $\mathcal{L}_S$ protocol can be written as $F(x)=x$.
On the other hand, 
if $\tau$ diverges, as in $\tau_L$, as the system size grows, the typical pair of nodes
becomes increasingly distant in the relative sense, and thus cannot in general remain in 
communication, or in a more formal
sense, protocols such as $\mathcal{L}_L$ may not be scalable~\cite{fn_Fx}. 

\begin{figure}[b]
\begin{center}
\includegraphics[width=8.cm]{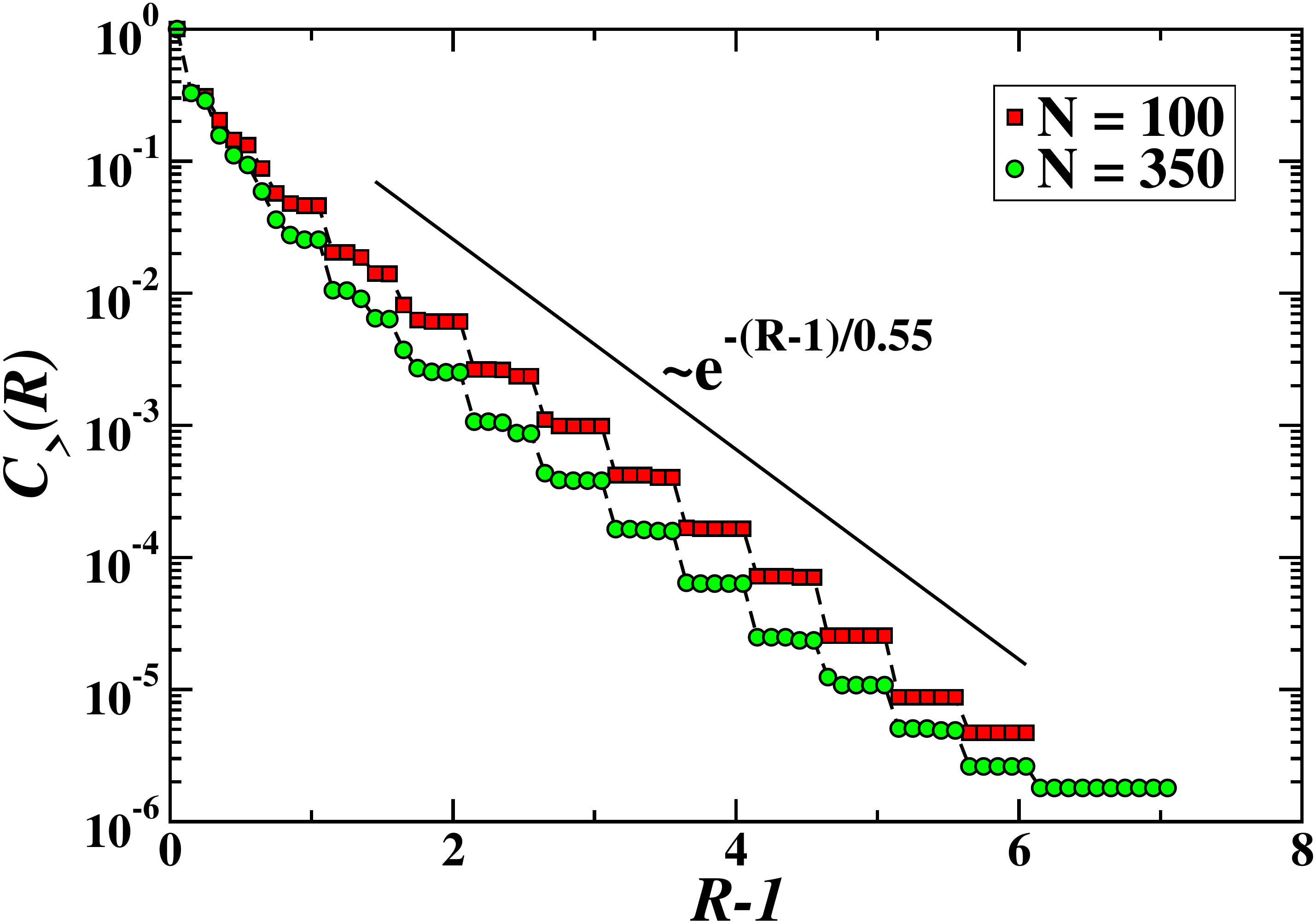}
\caption{Color Online) Cumulative distribution of the ratio between the length of the 
paths obtained with our protocol optimization method and those of the
corresponding original SP. The distribution becomes step-like due to the
discreteness of the length. }
\label{fig:6}
\end{center}
\end{figure}

The protocol we propose here has the behavior of $\mathcal{L}_S$ and thus, as in limited path percolation, 
can have a percolation threshold to a connected regime. 
Furthermore, the fact that we study scale-free degree distributed networks with small $\lambda$
indicates that from the standpoint of limited path percolation, there is always a connected regime.
In addition, when $\delta$ is close to 1, that the protocol is very robust as most pairs
of nodes remain connected after the introduction of the optimization of $b_{max}$.
These conclusions are also supported by results displayed in Figures~\ref{fig:4},~\ref{fig:5} and~\ref{fig:6}.

Furthermore, as explained above, a path obtained with our method can be only in two states:
either the SP configuration or, if they were expelled from a high used node and no length-degenerate alternative was available, in the shortest
configuration that does not pass through that node. Such
two state configuration recalls physical systems in which the components
can be in the energetic ground state or being activated by thermal (stochastic)
fluctuations to higher energy levels. To confirm whether this parallelism is
valid, we have plotted in Figure~\ref{fig:6} the cumulative distribution ($C_>(R)
= \int_R^\infty dR'\,P(R')$) of the ratios 
$R= \ell_{stat}/\ell_{SP}$, where $\ell_{stat}$ corresponds to the length of
each path in an stationary instance of the protocols obtained with our
optimization technique and $\ell_{SP}$ is the corresponding length of the
shortest path. As can be seen, the probability for the paths to be outside the
ground state (SP), in an "excited state", decays as an exponential 
function with a characteristic
ratio that depends on the graph (whether the graph is scale-free or not, and, if it is,
also on the exponent of the degree distribution). In the example of Figure~\ref{fig:6}, the
characteristic ratio is around $1.55$, 
which makes it extremely unlikely for any 
path to suffer a large stretching out of its SP length. This exponential decaying $C_>(R)$ resembles the Boltzmann factor in which $R-1$ plays the role of the energy.

\section{Weighted networks}
\label{weights}
The introduction of disorder in the network connections deeply modifies the previous scenario.
However, connections of divers quality can be  
expected in may real-world situations. A classical example within the 
context of transport is the Internet and the variability on bandwidth of
the connections between servers. Mathematically speaking, the quality of a 
connection can be represented by a scalar attached to each link that is known as 
its weight $w$. The presence of weighted links alters the definition of
shortest path \cite{park,newman01,newman04}. Instead of "short" in the sense of number 
of links between origin and destination, it may be important to find the path 
with the lowest weight along the way (or the largest, depending on the 
quality that the weight conveys). These paths are often referred to as 
optimal paths in the literature (see, for instance, \cite{cieplak}). 

Protocols based on Optimal Paths (OP) also have a very particular scaling 
with the network size~\cite{braunstein07}. Depending on the level of randomness on
the weights of the connections, the system can
fall into a strong- or a weak-disorder regime. The difference between these two
regimes is that in the strong-disorder case the fluctuations on the weight are
so large that the accumulated weight along each paths is
controlled by the edge with the highest weight, while in weak-disorder the "responsibility" for the path's overall weight is distributed among 
all the links along the path~\cite{wu,chen}. In the two regimes, the 
scaling of $b_{max}$, and of the average weight of the paths $\la w_{path} \ra$,
with $N$ changes with respect to unweighted graphs~\cite{goh05}. Here we will 
focus only on
the weak-disorder regime since it is the only one in which it makes sense to
search for an alternative protocol to the OP.  The cost of doing so in the
strong-disorder limit would diverge with the size of the graph.

One of the positive aspects of our protocol optimization is that its
generalization to weighted graphs is straightforward: at each step, a path
passing through
the node with the highest betweenness is chosen to be redirected. The path is 
then
exchanged by its alternative of lowest cost that does not pass through that 
node. As before, if there is no alternative, the path remains
invariant. In the following, we will study the performance of the stationary 
protocols produced by this method on
Molloy-Reed graphs with $\lambda = 2.5$ and with two possible functional forms
for the weight distributions: either an exponential 
$P(w) \sim e^{-w/w_c}$ or a power-law $P(w) \sim w^{-\beta}$. In order to
keep the results comparable for both distributions, we fix the parameter
$w_c$ to obtain a similar average weight as in the power-law distribution. For
the data shown, this will be $w_c = 2$ and $\beta = 2.5$. 

\begin{figure}
\begin{center}
\includegraphics[width=8.cm]{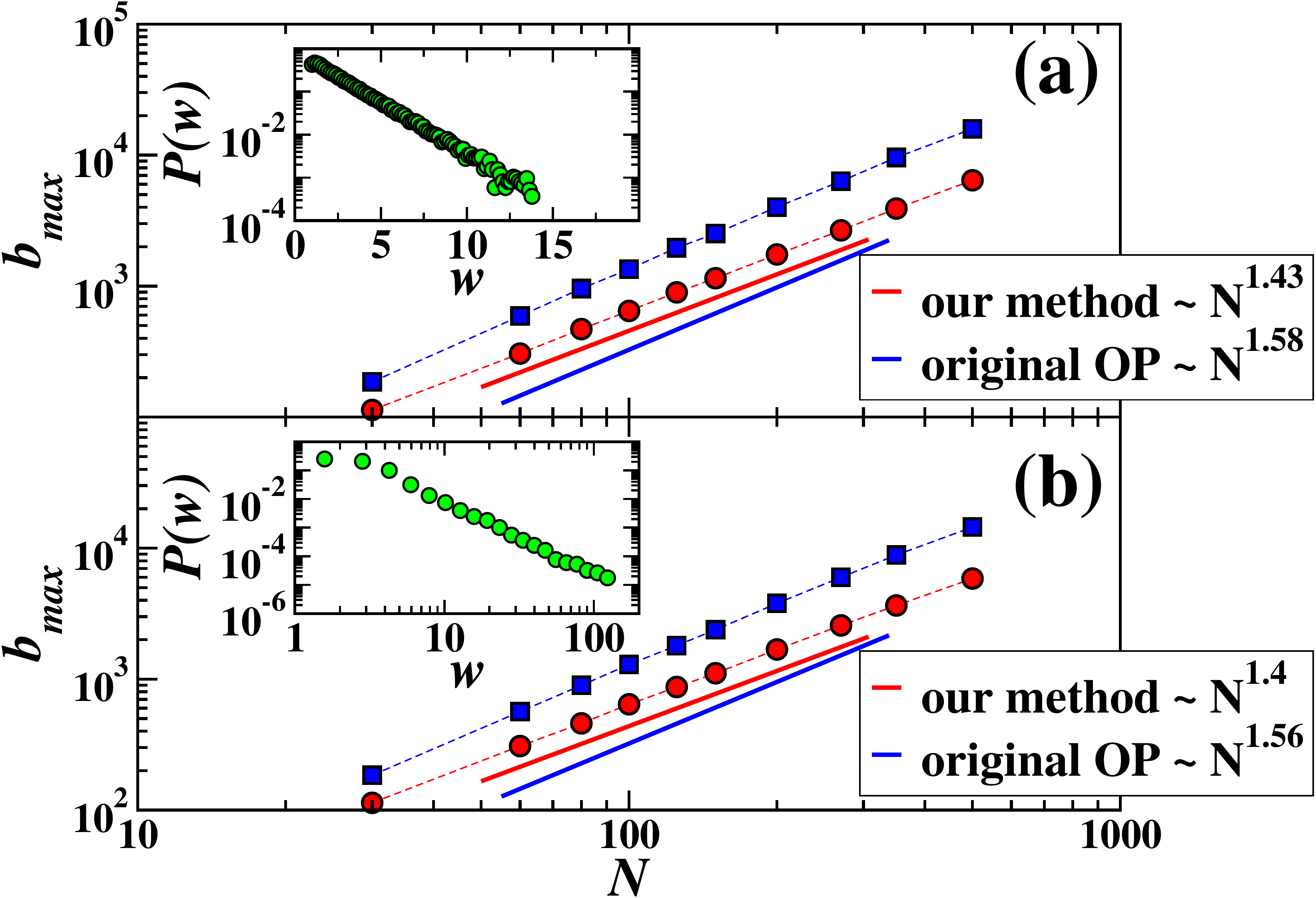}
\caption{(Color Online) Dependence of $b_{max}$ with the system size for 
weighted graphs. The (red) circles are the results for the protocols of our
method and the (blue) squares are for the original OP protocol. The weight
distribution in panel (a) follows an exponential $P(w) \sim e^{-w/2}$, and in (b)
the power-law $P(w) \sim w^{-2.5}$, as displayed in the insets for
networks of size $N = 500$.  The continuous curves in both panels are a guide to the eye showing  a visual comparison between the exponent values obtained by fitting the simulation results. }
\label{fig:7}
\end{center}
\end{figure}

\begin{figure}
\begin{center}
\includegraphics[width=8.cm]{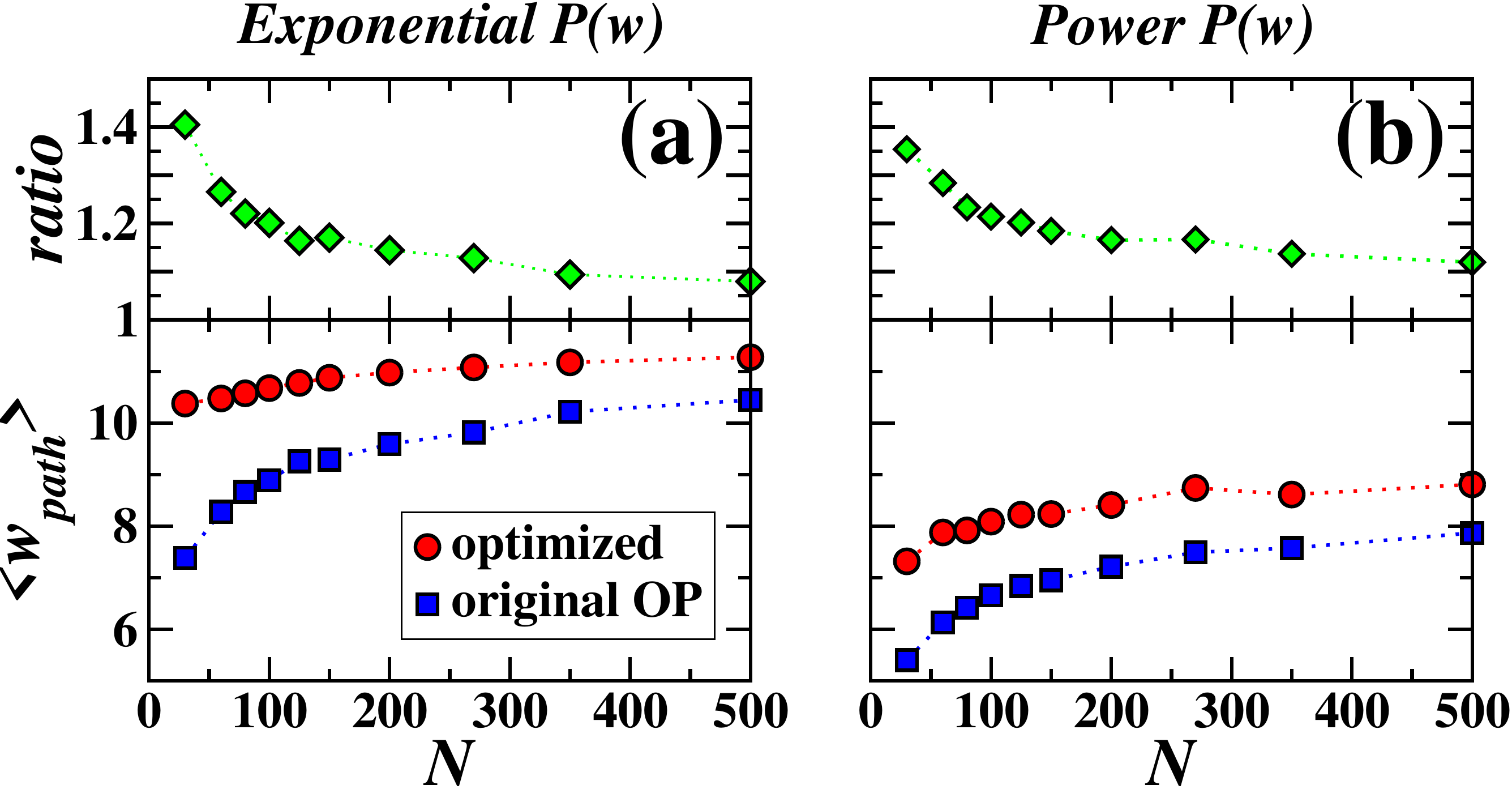}
\caption{(Color Online) Comparison between the average weight of the paths,
$\la w_{path} \ra$, in the original OP protocol and in the optimized protocols
with our method. The curves in (a) correspond to weighted graphs with an 
exponentially decaying weight distribution, while those in (b) have a 
power-law weight distribution.  }
\label{fig:8}
\end{center}
\end{figure}

The scaling of $b_{max}$ as a function of the network size is shown in Figure~\ref{fig:7}
for networks with both type of weight distributions. The application of the
optimization method produces a significant improvement in $b_{max}$ with
respect to the Optimal Paths protocol. For the range of system sizes of the
figure, this entails an improvement by a factor $3$ or even higher. Additionally, the exponent $\alpha$ of the
optimized protocols is close to $1.4$ while that of the original OP protocols
was about $1.55$. One can wonder about the price to pay for such an
improvement. In networks with continuous values of the weights, any
degeneracy in the weight of the alternative paths is extremely unlikely. 
Strictly speaking, then, there is always a price to pay, although it could be 
minute. To answer this question, we have plotted in Figure~\ref{fig:8} 
the 
average weight of the paths scaled with $N$ for protocols obtained with our
method and for the corresponding OP on both types of networks and both
weight distributions. Similarly to the unweighted case, we find
that the average weight is slightly higher but not extremely so. Also the ratio
between the average weight of the paths of our optimized protocols and of OP
decreases for increasing network sizes. Furthermore, even though it is not
shown in the figure, neither the shape of the distribution nor the weight of
the paths is hardly altered by the optimization of the protocols.

\section{Discussion and Conclusion}
\label{conclusions}

We study a routing protocol optimization method that takes advantage of the 
degeneracy of the shortest paths to reduce the congestion in the nodes of the 
network. The paths between node pairs in the optimized protocols can only be in 
two states: they can be either one of the shortest-paths or the closest 
alternative avoiding highly used nodes. We use ideas coming from Extremal Optimization 
to guide our algorithm, which proves to be an effective technique. The resulting
routing protocols show an appreciable reduction of the maximum node betweenness,
$b_{\rm max}$, while keeping the path-length distribution similar to the lower bound provided by the SP protocol. 
For the networks that we have analyzed $b_{\rm max}$ increases with $N$ as a 
power-law with an exponent $\alpha \approx 1.3$. That is only slightly higher than 
the best values reported in the literature, and, in contrast to those methods, the average path-length appears to
retain the $\log{(\log{(N)})}$-scaling of  the SP protocol. We show that methods which do not conserve a constant ratio
of path-length with the topological shortest paths end up in a disconnected
phase of percolation with limited paths, which is a possible indicator that 
on networks with packet loss such methods may not be scalable.

Our method is  easily generalized for weighted networks. We have studied two examples of weighted networks in the weak-disorder regime and found significant reductions in $b_{\rm max}$ with respect to the Optimal Paths protocol as well as in the scaling exponent with which $b_{\rm max}$ grows with $N$.

As we implemented the optimization method--a procedure in which we keep record 
of all the paths passing through each node--, the numerical performance 
is not optimal. Indeed, we were able to study a range of network sizes wide 
enough to characterize the scaling of the optimized protocols properties, but 
not as large as to allow for large network analysis. Our aim in this work was 
to explore the behavior of the optimization method and to ascertain whether 
the resulting routing protocols have interesting features concerning the size 
scaling and behavior of $b_{\rm max}$ and the path-length distributions. 
Indeed, the simple fact that we have shown that
it is feasible to perform such dual optimization is an important step forward.
Furthermore, we have found that this two parameter optimization is important because
without taking into account both $b_{max}$ and path-length, the protocols that
simply reduce $b_{max}$ may be less scalable. 

Therefore, the possibility of searching for efficient numerical methods to 
implement the optimization algorithm remains open for future work. A possible
solution, depending on the structure of the network, can be the implementation 
of a local search technique for the alternative paths. This local method could 
be highly efficient if the clustering (the density of triangles) in the graph is
high as typically occurs in geographically embedded or social driven networks. Locally tree-like structured networks, such as those considered here, typically require \emph{global} rearrangements between paths of nearly-degenerate length but hardly any overlap, as illustrated by Fig. 1a-b. Maintaining the paths of the routing protocol short can be crucial in situations
such as when the system suffers packet losses, a circumstance that seems to 
be relatively frequent in practice. Our optimization method represents in such 
cases a useful tool that offers a balance between betweenness reduction (lower 
congestion) and restricted path-lengths (reduced losses) in routing protocols.

\section{Acknowledgments}

EL acknowledges funding from the EPSRC grant EPSRC EP/E056997/1. JJR is funded by the FET grant 233847-Dynanets of the EU Commission. Also both JJR and SB acknowledge support by the U. S. National Science Foundation through grant
DMR-0312510.


\begin{thebibliography}{99}
\bibitem{stat} As a criterion to define the stationarity of the optimization process, we focus on 
the value of $b_{max}$ as a function of the iteration steps. $b_{max}$ is considered over the last 20 steps, and the process is said stationary if the 
relative error $E(b_{max}) \le 10^{-4}$. We confirmed the stationary of the betweenness and path-length distributions within numerical precision once this criterion is fulfilled. The number of path rearrangements required to 
achieve stationary protocols scales for the networks studied approximately as $\sim N^{2.5}$.
\bibitem{fn_Fx} Note that any function $F(x)$ scaling faster than $x$ and diverging is sufficient 
to obtain the same qualitative results. For instance, $F(x)=x\log x$ has the same effect. 
In general, if $\lim_{x\rightarrow\infty} F(x)/x\rightarrow\infty$
the paths scale too fast for a protocol satisfying limited paths to be scalable.
\end{thebibliography}
\end{document}